\documentclass[aps,prl,twocolumn,showpacs,superscriptaddress]{revtex4}
\usepackage{graphicx}

\begin{document}

\title{Phonon Spectra, Nearest Neighbors, and Mechanical Stability of Disordered Colloidal Clusters with Attractive Interactions}

\author{Peter J. Yunker}
\affiliation{Department of Physics and Astronomy, University of Pennsylvania, Philadelphia PA 19104, USA}
\author{Ke Chen}
\affiliation{Department of Physics and Astronomy, University of Pennsylvania, Philadelphia PA 19104, USA}
\author{Zexin Zhang}
\affiliation{Center for Soft Condensed Matter Physics and Interdisciplinary Research, Soochow University, Suzhou 215006, China}
\affiliation{Complex Assemblies of Soft Matter, CNRS-Rhodia-UPenn UMI 3254}
\author{A. G. Yodh}
\affiliation{Department of Physics and Astronomy, University of Pennsylvania, Philadelphia PA 19104, USA}

\date{\today}
\begin{abstract}
We investigate the influence of morphology and size on the vibrational properties of disordered clusters of colloidal particles with attractive interactions.  From measurements of displacement correlations between particles in each cluster, we extract vibrational properties of the corresponding ``shadow'' glassy cluster, with the same geometric configuration and interactions as the ``source" cluster but without damping.  Spectral features of the vibrational modes are found to depend strongly on the average number of nearest neighbors, $\overline{NN}$, but only weakly on the number of particles in each glassy cluster. In particular, the median phonon frequency, $\omega_{med}$, is essentially constant for $\overline{NN}$ $<2$ and then grows linearly with $\overline{NN}$ for $\overline{NN}$ $>2$. This behavior parallels concurrent observations about local isostatic structures, which are absent in clusters with $\overline{NN}$ $<2$ and then grow linearly in number for $\overline{NN}$$>2$.  Thus, cluster vibrational properties appear to be strongly connected to cluster mechanical stability (i.e., fraction of locally isostatic regions), and the scaling of $\omega_{med}$ with $\overline{NN}$ is reminiscent of the behavior of packings of spheres with repulsive interactions at the jamming transition.  Simulations of random networks of springs corroborate observations and suggest that connections between phonon spectra and nearest neighbor number are generic to disordered networks.
\end{abstract}

\pacs{63.50.Lm,61.43.Fs,63.22.Kn,64.70.kj,64.70.pv}
\maketitle
The phase behavior and vibrational properties of ensembles of \textit{repulsive} particles are determined largely by packing fraction [1].  Samples of monodisperse spheres, for example, gain structural order and eventually crystallize with increasing packing fraction [1], giving rise to low frequency plane-wave-like phonon modes.  In a related vein, ensembles of polydisperse spheres gain contacts with increased packing fraction, leading to vitrification [2] and ``soft phonon modes'' whose properties depend on average numbers of interparticle contacts [3].   By contrast to these ``space-filling'' systems, particles with strong \textit{attractive} interactions can form solid-like phases at low macroscopic packing fractions [4].  Dilute gels, for example, mechanically percolate across large distances [5], and disordered clusters containing relatively few particles often self-assemble into structures with large local packing fraction [6].  In this paper we explore how cluster morphology and cluster size affect the vibrational properties of disordered materials held together by strong attractive interactions. New understanding thus gained holds potential to elucidate fundamental differences between glassy materials composed of particles with attractive versus repulsive interactions, to uncover connections between vibrational spectra, mechanical stability, and the jamming problem, and to discover attributes of a disordered cluster that endow it with the properties of bulk glasses.

To date, a diverse collection of disordered systems have been observed to display surprising commonality in their vibrational properties.  Such systems include molecular [7], polymer [8], and colloidal glasses [9].  These disordered solids exhibit an excess of low frequency modes that are believed important for their mechanical and thermal properties [10]. The low frequency modes also appear connected to scaling and mechanical behaviors of \textit{repulsive} spheres near the zero-temperature jamming transition. At the jamming point, such disordered packings are ``isostatic'', i.e., they have exactly the number of contacts per particle required for mechanical stability; if a single contact is removed, the packing is no longer stable. Interestingly, marginal stability permits particle displacements that maintain isostaticity without energy cost; these motions are manifest as low frequency ``soft'' phonon modes [3,11].  When the sample packing fraction is increased above the jamming transition, the number of contacts per particle increases, the system is stabilized [12], and the number of soft modes is found to decrease [3].  In fact, the minimum soft mode frequency has been predicted to increase linearly with number of contacts per particle above the isostatic requirement [3].  Recent experiments have found some of these trends in thermal packings of repulsive particles [9], but application of such concepts to systems with attractive interactions has proven difficult. Ensembles of attractive particles can achieve isostaticity at arbitrary packing fraction, and even when they do not have enough contacts to be isostatic as a whole, the attractive systems can still have local mechanically stable regions [11]. Therefore, our study of vibrational properties in clusters of attractive particles will provide useful clues about underlying mechanisms responsible for the mechanical properties of disordered solids.

Herein we experimentally investigate the influence of cluster morphology and size on the vibrational properties of disordered clusters of colloidal particles with attractive interactions.  The disordered clusters with high local packing fractions are formed in water-lutidine suspensions where wetting effects induce fluid mediated attractions between micron-sized polystyrene particles. Each cluster is characterized by the number of particles it contains (N), an average number of nearest neighbors ($\overline{NN}$), and a number of local isostatic configurations (N$_{Iso}$). Displacement correlation matrix techniques employed in recent papers [9] are used to determine the phonon density of states of corresponding ``shadow'' attractive glass clusters with the same geometric configuration and interactions as the ``source'' experimental colloidal system but absent damping [9].  Surprisingly, the spectra and character of vibrational modes depend strongly on $\overline{NN}$ but only weakly on N. The median phonon frequency, $\omega_{med}$, which characterizes the distribution of low and high frequency modes, is observed to be essentially constant for $\overline{NN}$ $<2$, and then grows linearly with $\overline{NN}$ for $\overline{NN}$ $>2$. This behavior parallels concurrent observations about local isostatic structures, which are absent in clusters with $\overline{NN}$ $<2$ and then grow linearly in number for $\overline{NN}$ $>2$. Thus cluster vibrational properties appear to be strongly connected to cluster mechanical stability (i.e., fraction of locally isostatic regions), and the scaling of $\omega_{med}$ with $\overline{NN}$ is reminiscent of the behavior of packings of spheres with repulsive interactions at the jamming transition.  Simulations of random networks of springs corroborate observations and further suggest that connections between phonon spectra and nearest neighbor number are generic to disordered networks.

\begin{figure}
\scalebox{0.32}{\includegraphics{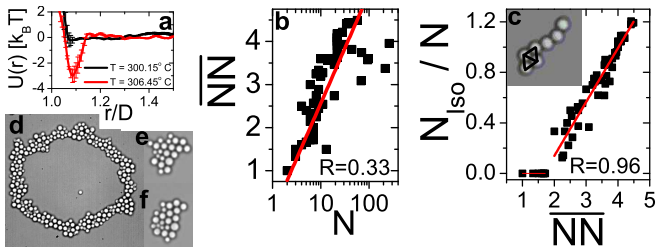}}
\caption{ a. Plot of the temperature-dependent (T$ = 300.15$ K and $306.45$ K) interparticle potential, u(r), as a function of particle separation (r) normalized by particle diameter (D). The temperature dependence is induced by wetting effects in near-critical water-lutidine mixtures.  b. The number-fraction of locally isostatic configurations per particle, N$_{Iso}/$N is plotted versus average number of nearest neighbors, $\overline{NN}$. Solid lines are linear fits within two separate regimes. c. Plot summarizing N and $\overline{NN}$ in every cluster. d-f. Experimental snapshots of clusters with N$= 261$, $\overline{NN}$$= 3.95$ (d), N$= 22$, $\overline{NN}$$= 3.91$ (e), and N$ = 22$, $\overline{NN}$$ = 4.09$ (f).
\label{exp_info}}
\end{figure}

The experiments employ bidisperse suspensions of micron-sized polystyrene particles (Invitrogen), with diameters $d_{S} = 1.5 \mu$m and $d_{L} = 1.9 \mu$m, and number ratio 1:2, respectively. Binary mixtures of particles were used to minimize crystallization effects.  Particles were suspended in a mixture of water and 2, 6-lutidine near its critical composition, i.e., with lutidine mass fraction of $0.28$.  Colloidal particles suspended in this near-critical binary mixture, experience temperature dependent fluid-mediated repulsive or attractive interactions [13]. Interparticle potentials were determined from  measurements of the particle pair correlation function with  liquid structure theory and image artifact corrections [14] (Fig.\ 1a).  Many different disordered particle clusters are created by first suspending particles deep in the repulsive regime (300.15 K), and then increasing the sample temperature (to $306.5 K$) \textit{in situ}.  Sample temperature control was accomplished using an objective heater (Bioptechs) connected to the microscope oil immersion objective [15]. Particles are confined between two glass coverslips (Fisher) with a spacing of $\sim$$(1.1 \pm 0.05) d_{L}$, making the sample system quasi-2D. The glass cell was treated with NaOH, so the particle-wall interaction potential is repulsive at relevant temperatures [16].  The global area fraction is $\sim$$0.2$.  Disordered clusters of various sizes and shapes self-assemble. Other clusters are assembled with aid of laser tweezers [17], either by grabbing particles and adding them to existing clusters, or by dragging an optical trap across a cluster and forcing rearrangements.  Samples equilibrated for about six hours, and video data were collected at a rate of $10$ frames per second after equilibration for $\sim1000$ seconds.

As noted above, the particle cluster structure is characterized by several factors, including average number of nearest neighbors per particle and number of locally isostatic configurations. Neighbors are defined as particles spatially separated by less than a cutoff distance, a distance equal to the first minimum in the particle pair correlation function.  Local isostatic regions consist of three particles ($a$, $b$, and $c$) that are mutually nearest neighbors (i.e., $a$ and $b$ are neighbors, $a$ and $c$ are neighbors, and $b$ and $c$ are neighbors). Plots summarizing N, $\overline{NN}$, and N$_{Iso}$ for each cluster studied are shown in Fig.\ 1b and c, along with experimental snapshots of selected clusters (Fig.\ 1d-f). Note that $\overline{NN}$ tends to increase with increasing N for our distribution of cluster sizes, but that the increase is not monotonic. The dependence of N$_{Iso}$ on $\overline{NN}$ exhibits two regimes. Specifically, N$_{Iso}/N$ is $0$ for $\overline{NN}$$<2$, becomes non-zero abruptly at $\overline{NN}$$=2$, and then grows linearly with $\overline{NN}$ for $\overline{NN}$$>2$.  Thus, we identify $\overline{NN}$$=2$ as the ``local isostatic'' point.

\begin{figure*}[!t]
\includegraphics[width=\textwidth]{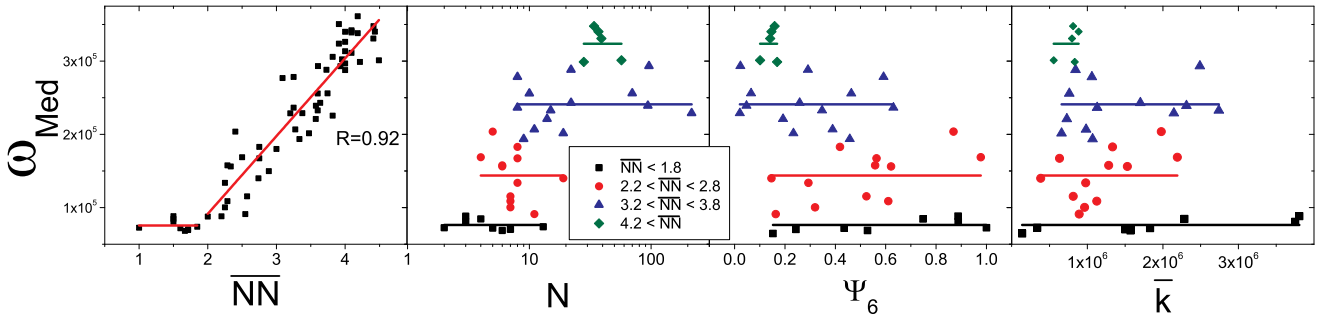}
\caption{ a. Median frequency, $\omega_{med}$, versus average number of nearest neighbors, $\overline{NN}$. Two regimes exist. For $\overline{NN}$$<$$2$, $\omega_{med}$ is constant (line is constant fit). For $\overline{NN}$$>$$2$, $\omega_{med}$ increases linearly with $\overline{NN}$ (line is a linear fit). In b-d, $\overline{NN}$ in different sample-types is indicated by plotting $\overline{NN}$$<$$1.8$ as black squares, $2.2$$<$$\overline{NN}$$<$$2.8$ as red circles, $3.2$$<$$\overline{NN}$$<$$3.8$ as blue triangles, $4.2$$<$$\overline{NN}$ as green diamonds. Lines are best constant fits. b. $\omega_{med}$ versus number of particles, N. c. $\omega_{med}$ versus orientational order parameter, $\psi_{6}$. d. $\omega_{med}$ versus average total nearest neighbor spring constant, $k$.  \label{fig:modes}}
\end{figure*}

The vibrational properties of each cluster are extracted by measuring displacement correlations of the particles within each cluster [9].  Briefly, we extract the time-averaged displacement correlation matrix, which, in the harmonic approximation, is directly related to the cluster stiffness matrix, defined as the matrix of second derivatives of the effective pair interaction potential with respect to particle position displacements.  These measurements of the stiffness matrix permit us to construct and derive properties of a ``shadow'' glass system that has the same static properties as our colloidal system [9].  Following [18], we expect undamped particles that repel at very short-range, and attract due to fluid mediated effects, to give rise to solid-like vibrational behavior on time scales long compared to particle collision times but short compared to the time between particle rearrangement events [9].  Data were collected over $1000$ seconds so that the number of degrees of freedom, $8 \leq 2N \leq 500$, is small compared to the number of time frames ($> 10\times2N $) [9].  Additionally, we find elements of the stiffness matrix to be far above the noise only for adjacent particles, as expected.

Comparing the frequency spectra of clusters with small N can be challenging, since not enough modes are present for clear identification of a traditional ``peak'' frequency, and since fluctuations can significantly shift mode frequency.  For these reasons we choose to characterize each cluster's density of states by its median frequency, $\omega_{med}$, i.e., we choose the frequency, $\omega_{med}$, such that half of the cluster mode frequencies are smaller than $\omega_{med}$ and half are larger.  Plots of $\omega_{med}$ as a function of average number of nearest neighbors, $\overline{NN}$, and as a function of total number of cluster particles, N (at nearly fixed $\overline{NN}$), are shown in Fig. 2a and b.

Surprisingly, $\omega_{med}$ has little correlation with N, exhibiting linear correlation coefficients of $R=0.29$.  This effect is even more apparent when the number of nearest neighbors is held nearly constant (Fig.\ 2b).  However, $\omega_{med}$ depends strongly on the average number of nearest neighbors ($\overline{NN}$).  We observe two distinct regimes in this case.  For $\overline{NN}$$<$$2$, $\omega_{med}$ is constant. For $\overline{NN}$$>$$2$, $\omega_{med}$ increases linearly with $\overline{NN}$ ($R=0.92$). Interestingly, the dependence of $\omega_{med}$ on $\overline{NN}$ is very similar to the dependence of the number-fraction of locally isostatic configurations per particle, i.e., N$_{Iso}$/N, on $\overline{NN}$ (Fig.\ 1b).  These observations suggest that the vibrational properties of disordered clusters are strongly dependent on the presence of locally rigid elements. Note, we also expect to observe a correlation between $\omega_{med}$ and N for our cluster distribution; this correlation arises because $\overline{NN}$ tends to increase with N for typical cluster distributions.  Thus, while we expect the vibrational spectra of a disordered attractive cluster to become similar to that of a bulk glass as the total number of particles in the cluster increases, the underlying mechanism for this effect originates from the average number of nearest neighbors in the cluster, rather than the total particle number in the cluster.

Interestingly, the dependence of $\omega_{med}$ on $\overline{NN}$ is also reminiscent of the behavior of hard-spheres in the vicinity of the zero-temperature jamming transition [4]. In this case, the characteristic frequency, $\omega^{*}$, of excess quasi-localized or ``soft'' modes increases linearly with $\overline{NN}$ when $\overline{NN}$$>$$\overline{NN}$$_{C}$, where $\overline{NN}$$_{C}$ is the number of contacts necessary for isostaticity. Similarly, in our experiments with attractive particles, $\omega_{med}$ increases linearly with $\overline{NN}$ when $\overline{NN}$$>$$2$ and when locally rigid elements are present. In thermal experiments with \textit{repulsive particles}, $\omega_{med}$ shows a strong linear correlation with $\omega^{*}$ ($R=$$0.96$), and $\omega_{med}$ has a strong linear relationship with $\overline{NN}$ [9].  Thus our observations suggest that similar ``jamming transition'' physics may control properties of both highly packed ``repulsive'' glasses, composed of particles with repulsive interactions, and disordered clusters composed of particles with attractive interactions low (overall) packing fraction.

As a final demonstration of the importance of average-number-of-nearest-neighbors versus total-number-of-particles in a cluster, consider two clusters that look very different (Fig.\ 1 b and c) but have almost the same number of average nearest neighbors ($\overline{NN}$).  These clusters have similar characteristic frequencies (i.e., $\omega_{med}= 3.0 \times 10^{5}$ and $3.1 \times 10^{5}$ for Fig.\ 1b and c, respectively).  On the other hand, two clusters that contain the exact same number of particles but have different $\overline{NN}$ (Fig.\ c and d) possess a set of very different characteristic frequencies ($\omega_{med}= 3.1 \times 10^{5}$ and $3.6 \times 10^{5}$ for Fig.\ 1c and d, respectively).

As per other calculable cluster properties, $\omega_{med}$ does not appear to correlate strongly with many traditional structural quantities. The bond orientational order parameter, $\psi_{6}$ [19], for example, does not correlate strongly with $\omega_{med}$, when $\overline{NN}$ is held approximately constant ($R=0.44$) (Figure 2). Similarly, the average stiffness between nearest neighbor pairs, $\bar{k}$ [20], does not correlate substantially with $\omega_{med}$, when $\overline{NN}$ is held approximately constant ($R=0.10$).  Thus, little correlation exists between local order and $\omega_{med}$.

\begin{figure}
\scalebox{0.32}{\includegraphics{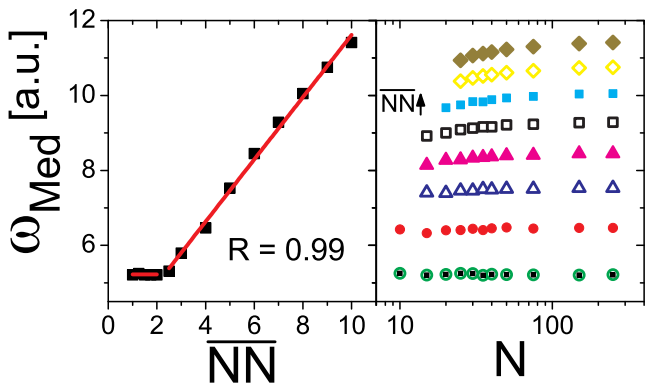}}
\caption{ a. Median frequency, $\omega_{med}$, plotted versus average number of nearest neighbors, $\overline{NN}$, from simulations of random matrices. b. Median frequency, $\omega_{med}$, plotted versus number of particles, N, from random matrices with $\overline{NN}$$ = 1$ (closed squares), $2$ (open circles), $3$ (solid circles), $4$ (open triangles), $5$ (solid triangles), $6$ (open squares), $7$ (solid squares), $8$ (open diamonds), and $9$ (solid diamonds). \label{exp_info}}
\end{figure}

Unfortunately, in experimental systems, structural parameters cannot be tuned independently.  For example, clusters with large N and small $\overline{NN}$ are difficult to find or create. Therefore, as a final check on the importance of structural quantities other than $\overline{NN}$, we explored the calculated spectra of randomly generated networks of springs. Random networks of springs, expressed as matrices, $K_{ij}$, were generated following a few simple rules that ensure the matrices only contain information about N and $\overline{NN}$ [21]: (1) Each element, $ij$, in the matrix represents the spring constant between particle/coordinate $i$ and particle/coordinate $j$; (2) the number of rows/columns in these symmetric matrices is twice the number of particles, representing each coordinate of each particle; (3) the number of off diagonal elements greater than zero is equivalent to the number of nearest neighbors (each particle has the same number of nonzero elements); (4) diagonal elements are set such that the sum of each row/column is zero, ensuring translational invariance.  Thus N and $\overline{NN}$ can be varied independently. For every combination of N and $\overline{NN}$, $10,000$ random matrices are generated. $\omega_{med}$ is calculated from the combination of all generated frequencies (Figure 3). Note, many of these networks cannot be duplicated in real experimental systems, since nearest neighbor pairings are assigned at random and not based on proximity. Nevertheless, we found that $\omega_{med}$ follows the same trends in these simulations as observed in our experiments: $\omega_{med}$ has little or no correlation with N (i.e., with $\overline{NN}$ held constant, $\omega_{med}$ changes by less than $5\%$), but exhibits a strong correlation with $\overline{NN}$ ($R> 0.99$). Thus $\omega_{med}$, a dominant feature of cluster vibrational spectra, appears to be the result of network connectivity, rather than a result of specific structure.

In conclusion, the spectra and character of vibrational modes in disordered ``attractive'' clusters do not depend strongly on the number of particles in the cluster, but do depend strongly on the average number of nearest neighbors and the number of locally isostatic configurations. The fact that $\omega_{med}$ depends on $\overline{NN}$, but not on total number and packing fraction, suggests that these disordered clusters are a useful model system for network glasses (e.g., silica [22]). Network glasses are composed of particles (e.g., molecules) that have directional bond forming interactions which set $\overline{NN}$ [23], leading to the formation of solids at low packing fractions. In fact, the vibrational [24] and mechanical [25] properties of network glasses are believed to depend strongly on $\overline{NN}$. Thus, the disordered clusters composed of particles with attractive interactions that we have introduced could serve as a convenient model system for network glasses and their many applications (e.g., non-crystalline semiconductors [26]).\\

\begin{acknowledgments}
We thank Piotr Habdas, Carl Goodrich, Andrea Liu, and Tim Still for helpful discussions. We gratefully acknowledge financial support from the National Science Foundation through DMR-0804881, the PENN MRSEC DMR-0520020, and NASA NNX08AO0G.
\end{acknowledgments}
\vspace{\baselineskip}
\noindent[1] W. G. Hoover and F. H. Ree, J. Chem. Phys. \textbf{49}, 3609 (1968); P. N. Pusey and W. van Megen, Nature \textbf{320}, 340 (1986).\newline
[2] A. van Blaaderen and P. Wiltzius, Science \textbf{270}, 1177 (1995).\newline
[3] M. Wyart, S. R. Nagel, and T. A. Witten, Euro. Phys. Lett. \textbf{72}, 486 (2005).\newline
[4] E. Zaccarelli, J. Phys. Condens. Matter \textbf{19}, 323101 (2007).\newline
[5] M. Y. Lin, \textit{et al.}, Nature \textbf{339}, 360 (1989).\newline
[6] G. Meng, \textit{et al.}, Science \textbf{327}, 560 (2010); P. J. Lu, \textit{et al.}, Phys. Rev. Lett. 96, 028306 (2006).\newline
[7] A. P. Sokolov, \textit{et al.}, Phys. Rev. Lett. \textbf{69}, 1540 (1992).\newline
[8] B. Frick and D. Richter, Science 267, 1939 (1995).\newline
[9] D. Kaya, \textit{et al.}, Science \textbf{329}, 656 (2010); A. Ghosh, \textit{et al.}, Phys. Rev. Lett. \textbf{104}, 248305 (2010); K. Chen, \textit{et al.}, \textbf{105}, 025501 Phys. Rev. Lett. (2010).\newline
[10] R. O. Pohl, X. Liu, and E. Thompson, Rev. Mod. Phys. \textbf{74}, 991 (2002).\newline
[11] D. J. Jacobs and M. F. Thorpe, Phys. Rev. Lett. \textbf{75}, 4051 (1995).\newline
[12] C. S. O'Hern, \textit{et al.}, Phys. Rev. Lett. \textbf{88}, 075507 (2002).\newline
[13] D. Beysens and T. Narayanan, J. Stat. Phys. \textbf{95}, 997 (1999); C. Hertlein, \textit{et al.}, Nature \textbf{451}, 172 (2008).\newline
[14] Y. L. Han, and D. G. Grier, Physical Review Letters 91, 038302 (2003).
[15] Z. Zhang, \textit{et al.}, Nature \textbf{459}, 230 (2009); P. Yunker, \textit{et al.}, Phys. Rev. Lett. \textbf{103}, 115701 (2009); P. Yunker, Z. Zhang, and A. G. Yodh, Phys. Rev. Lett. \textbf{104}, 015701 (2010).\newline
[16] F. Soyka, \textit{et al.}, Phys. Rev. Lett. \textbf{101}, 208301 (2008).\newline
[17] D. G. Grier, Nature \textbf{424}, 810 (2003).\newline
[18] C. Brito and M. Wyart, Euro. Phys. Lett. \textbf{76}, 149 (2006).\newline
[19] $\psi_{6}=1/N \sum_{j=1..N}\sum_{k=1..NN_{j}}exp(i\theta_{jk}\pi/3)/NN_{j}$, where $\theta_{jk}$ is the angle between particles $j$ and $k$, and $NN_{j}$ is the number of neighbors for particle $j$.\newline
[20] $\bar{k}=1/N\sum_{i=1..N}\sum_{j=1..NN_{i}}|K_{ij}|/NN_{i}$.\newline
[21] M. Wyart, Euro. Phys. Lett. \textbf{89}, 64001 (2010); V. Gurarie and J. T. Chalker, Phys. Rev. Lett. \textbf{89}, 136801 (2002); Y. M. Beltukov and D. A. Parshin arxiv.org/abs/1011.2955 (2010).\newline
[22] P. H. Gaskell and D. J. Wallis, Phys. Rev. Lett. Letters \textbf{76}, 66 (1996).\newline
[23] F. Sciortino, Euro. Phys. J. B \textbf{64}, 505 (2008).\newline
[24] W. A. Kamitakahara, \textit{et al.}, Phys. Rev. B \textbf{44}, 94 (1991).\newline
[25] M. Zhang, \textit{et al.}, J. Non-Cryst. Solids \textbf{151}, 149 (1992).\newline
[26] T. Kamiya, K. Nomura, and H. Hosono, Sci. and Tech. of Adv. Mater. \textbf{11}, 044305 (2010).\newline

\end{document}